\documentclass[twocolumn]{aastex701}
\hypersetup{linkcolor=red,filecolor=cyan,urlcolor=magenta}

\usepackage{xcolor}

\newcommand{\msun}{M$_{\odot}$}

\newcommand{\kms}{km s$^{-1}$}

\newcommand{\vesc}{$v_{\rm esc}$}

\newcommand{\SB}{\textcolor{red}{SB24}}
\newcommand{\qin}{$q_{\rm in}$}
\newcommand{\qout}{$q_{\rm out}$}

\usepackage{xcolor}
\usepackage{lipsum}
\usepackage[normalem]{ulem}
\usepackage{amsmath}

\begin{document}

\title{Mergers and Recoil in Triple Massive Black Hole Systems from Illustris}

\author{Pranav Satheesh}
\email[show]{pranavsatheesh@ufl.edu}
\affiliation{Department of Physics, University of Florida, Gainesville, FL 32611, USA}

\author{Laura Blecha}
\email{lblecha@ufl.edu}
\affiliation{Department of Physics, University of Florida, Gainesville, FL 32611, USA}

\author{Luke Zoltan Kelley}
\email{lzkelley@berkeley.edu}
\affiliation{Department of Astronomy,University of California at Berkeley, Berkeley, CA 94720, USA}

\begin{abstract}
Massive black hole binaries (MBHBs) form through galaxy mergers and are among the loudest sources of gravitational waves (GWs) in the universe. If the binary inspiral time is long, a subsequent galaxy merger can introduce a third black hole, forming a triple system. In the Illustris cosmological simulation, 6\% of MBHBs form such triples at parsec scales, where strong three-body interactions are likely. We apply results from numerical simulations of triple MBHs to strong triples identified in Illustris to assess their impact on MBH mergers and recoils. We find that strong triple interactions increase the overall merger fraction by $4\%$. Including triple interactions raises the merger fraction of MBHs in strong triple systems from 40\% to 69\%, relative to modeling binary evolution in isolation. Furthermore, massive, major mergers are over three times more likely to be facilitated by strong triple interactions than mergers in general. We also compare GW recoil kicks to gravitational slingshot kicks from triple interactions. Both mechanisms can produce kicks exceeding host escape speeds, ejecting MBHs and producing wandering or offset black holes. Although slingshots yield the highest velocity kicks, GW recoils dominate the ejected population when assuming random MBH spin orientations. Under this assumption, ejections from GW recoil and slingshot kicks reduce the total number of mergers by 6\%. Our results highlight the impact of strong triple dynamics and GW recoils on MBH evolution and support their inclusion in cosmological simulations.
\end{abstract}


\section{Introduction}
\label{sec:intro}
Massive black holes occupy the centers of most galaxies \citep{Kormendy_1995,magorrian_demography_1998}, with observational evidence pointing to correlations between the MBH mass and several properties of the host's stellar bulge \citep{ferrarese_fundamental_2000,tremaine_slope_2002,gultekin_m-sigma_2009,mcconnell_revisiting_2013,Kormendy_2013}. This relationship implies that the growth of MBHs is intertwined with galaxy evolution. In the hierarchical model of galaxy evolution, mergers are considered a crucial component. Since MBHs are commonly found in galactic centers, an MBH binary (MBHB) can form as a natural consequence of galaxy mergers \citep{Begelman1980}.

As these MBHs dynamically inspiral closer together by interactions with stars and gas, they may eventually reach sub-parsec (sub-pc) orbital separations. At $\sim$ mpc scales,  they inspiral via the emission of gravitational waves (GWs) \citep{PetersandMathews1964}. MBHBs are the loudest GW sources in the universe, with chirp frequencies ranging from millihertz (mHz) for MBHs around $\sim 10^6$ \msun{} to nanohertz (nHz) for MBHs around $10^9$ \msun{} \citep{Sesana2013}. In the nanohertz frequency range, GWs from a population of MBHBs can combine to produce a stochastic gravitational wave background (GWB). Pulsar timing arrays (PTAs) around the world have found compelling evidence for a GWB that is consistent with an MBHB origin of the background \citep{agazie_nanograv_2023,antoniadis_second_2023,reardon_search_2023,xu_searching_2023}. The future Laser Interferometer Space Antenna (LISA) will be able to see MBHB mergers for masses  $\lesssim 10^6$  \msun{} out to a redshift of $z \sim 20$ \citep{amaroseoane2017laserinterferometerspaceantenna}. 

To accurately predict the GW signatures from the merging MBHB population, we need to understand the timescales of their mergers, which remain highly uncertain. The MBHB evolutionary timescale is typically of the order of Gyrs and previous works using cosmological simulations \citep{Blecha2016,Salcido_2016} have used a fixed time delay of between a galaxy merger and MBHB coalescence. \citep{Salcido_2016} uses a fixed delay of 0.1 Gyr for gas-rich galaxies and 5 Gyr for gas-poor galaxies. In contrast, works that model binary evolution physics in postprocessing \citep{Kelley_2017a,Katz2020,Volonteri_2020} find that many binaries stall and fail to merge within a Hubble time. For example, the fiducial model in \citet{Kelley_2017a} sees a peak of MBHB lifetimes at a median value is 29 Gyr, with only $\sim 7$ \% binaries have lifetime less than 1 Gyr.

MBH binary inspiral or ``binary hardening" is driven by various physical processes first detailed in the work by \citet{Begelman1980}. Following a galaxy merger, dynamical friction \citep{Chandreshkar1943,Antonini2012} is responsible for bringing the MBHs towards the galactic center, leading to the formation of a gravitationally bound binary. The MBHB becomes a gravitationally bound system when the mass enclosed by the binary orbit is comparable to the MBH mass, which typically happens around $\lesssim$ 1-10 pc \citep{Begelman1980,quinlan_dynamical_1996,Yu_2002}. Once the binary becomes bound, stars with sufficiently low angular momentum, occupying the loss cone (LC), will scatter off the binary and continue to harden it \citep{sesana2008, quinlan_dynamical_1996,quinlan_dynamical_1997, Merritt_review_2005}. At sub-pc scales, in gas-rich mergers, binary hardening can happen via interactions with a circumbinary disk (CD) \citep{Dotti2010,cuadra_massive_2009,nixon_tearing_2013,goicovic_infalling_2017,siwek_orbital_2023,siwek_preferential_2023,siwek_signatures_2024}. The depletion of stars in the LC can cause the binary to stall at pc scales, known as the ``final parsec problem" \citep{milosavljevic_final_2003}. However, the LC can be efficiently refilled in several cases \citep{Yu_2002, holleybockelmann2006lossconetriaxialgalaxies,berczik_2006,Holley_Bockelmann_2010,Preto_2011,Khan_2011,Holley-Bockelmann2015,Khan_2016}, overcoming the stalling.

Another process that can aid in merging MBHs, particularly if binaries are stalled due to inefficient LC replenishment, and in gas-poor
environments, involves interactions with a third MBH \citep[e.g.,][]{hoffman_dynamics_2007,ryu_interactions_2017,bonetti_post-newtonian_2018-1}. If the binary inspiral time is long enough to allow a new galaxy merger to bring a third, ``intruder" BH close, then a triple system can form. In hierarchical triple systems, Kozai-Lidov (K-L) oscillations \citep{Kozai1962,lidov_evolution_1962,Naoz_2016} can secularly increase the orbital eccentricity of the inner binary, driving it to coalescence. If the intruder reaches the galactic nucleus at distances comparable to the inner binary’s orbital separation, chaotic three-body interactions can trigger a prompt MBH merger. These interactions also frequently result in the ejection of the lightest BH via gravitational slingshot, leaving the more massive pair tightly bound and able to merge on shorter timescales \citep{saslaw_gravitational_1974,hills_encounters_1975,blaes_kozai_2002,iwasawa_evolution_2006,hoffman_dynamics_2007}.

Previous numerical studies of such as \citet{hoffman_dynamics_2007} studied MBH triplets in massive galaxies and showed that three-body interactions enhance the MBHB coalescence rate. They also showed that triple interactions lead to a population of highly-eccentric binaries and also produce a population of ``wandering" black holes due to slingshot ejections. Similarly, \citet{volonteri_assembly_2003} tracked the formation of triple BHs in halo merger trees and, using a simple prescription of triple interactions, found that slingshot ejections may create a population of wandering black holes.

Although \citet{hoffman_dynamics_2007} used a Newtonian treatment of the dynamics, major improvements were made by \citet{bonetti_post-newtonian_2016,bonetti_post-newtonian_2018-1}, who included relativistic corrections up to 2.5 Post-Newtonian (PN) order in their triple MBH dynamics code and
explored a wide parameter space of triple systems. In a follow-up paper \cite{bonetti_post-newtonian_2018}, they use a semi-analytical model (SAM) of galaxy and massive black hole evolution with triple interaction included by interpolating the numerical simulation results from \citet{bonetti_post-newtonian_2018-1}. 
They found that even if all MBH binaries stall, triple encounters could still produce an observable GWB for PTAs. Moreover, in a follow-up study, \citet{bonetti_post-newtonian_2019} showed that triple interactions might also significantly contribute to LISA-detectable events. 

Because triple interactions can eject the lightest BH via a gravitational slingshot, they may produce
offset or wandering MBHs, which, in some cases, could manifest as observable offset active galactic nuclei (AGN) \citep{Madau_2004,Loeb_2007, blecha_effects_2008, barrows_spatially_2016, Blecha_2019_Astro2020}. \citet{vanDokkum2023} recently identified a slingshot recoil candidate, but subsequent studies have
favored a bulgeless edge-on galaxy explanation \citep{Sanchez2023,Sanchez2023b,Montes2024}.

BH recoil kicks can also occur following a MBH merger: when merging binaries have unequal masses or spins, they emit asymmetric GW 
radiation, resulting in a GW recoil that displaces the newly merged MBH \citep[e.g.,][]{bekenstein_gravitational-radiation_1973, Campanelli_2007}. Numerical relativity (NR) simulations have shown that GW recoils may reach $\sim$ 5000 \kms \citep{Campanelli_2007,lousto_orbital_2011}. Kicks of even a few hundred \kms\ can eject a MBH from its host galaxy center \citep{Lousto_2012, Gerosa_2014, Schnittman_2007, Ricarte2021} and can potentially impact MBH merger rates  \citep{Sesana_2009}. If the MBH is actively accreting at the time of the kick, it will carry along its accretion disk and broad-line region and may appear as an off-center AGN \citep{Loeb_2007,blecha_effects_2008, Volonteri_2008, Komossa_2012}. A handful of  recoiling AGN candidates have been observed \citep{Komossa2008,Civano2010,Koss2014,Kalfountzou2017,Chiaberge_2017,Hogg_2021,Ward2021,vanDokkum2023,bigmac2024,Uppal_2024}. Conclusive determination of the nature of these candidates has proven difficult, but prominent candidates have been ruled out \citep{Decarli2014,Li2024} and a confirmed recoil has recently been claimed for another candidate \citep{chiaberge2025}. 

To understand the relative contributions of slingshot kicks and GW recoils in producing wandering and offset BHs, it is important to characterize triple systems and their outcomes. \citet{sayeb_mbh_2023} identified triple MBH populations after applying the binary inspiral model \citet{Kelley_2017a} to the MBH binaries in the Illustris cosmological simulation in post-processing. They identified 
instances where an intruder MBH overtakes a binary, forming a triple system. Their results show that in their fiducial model,
22 \% of the binaries form a triple system, with over $> 70 \%$ being binaries that would not otherwise merge by $z=0$. Additionally, they find that $\sim 6 \%$ of the binaries form ``strong" triples at $\sim$ pc scale separations. 

This work builds on \citet{sayeb_mbh_2023} by incorporating triple MBH simulation results from \citet{bonetti_post-newtonian_2018-1} to analyze the outcomes of systems in Illustris undergoing strong triple interactions. This approach is similar in spirit to SAMs \citep{bonetti_post-newtonian_2018,bonetti_post-newtonian_2019,Izquierdo_Villalba_2021} that include the same subgrid triple interaction model. The primary goal of this work is to evaluate the impact of strong triple interactions on the MBHB population in Illustris, including the impact of triples on MBH mergers as well as recoiling MBHs produced by both GW recoil and slingshot kicks.

This paper is structured as follows. In Section \ref{sec: methods} we describe the MBH binary and triple population, the hardening mechanisms and the method used in calculating the recoil velocities. In Section \ref{sec: results} we go over our key results and finally in Section \ref{sec: discussion} we discuss our conclusions. 

\section{Methods}
\label{sec: methods}

For our work, we use data from the Illustris cosmological hydrodynamic simulation suite \citep{Vogelsberger_2013,Genel_2014,Nelson_2015}. Since Illustris cannot probe dynamics below the scale of the gravitational softening length defined for each particle or cell, the dynamics of the binary inspiral need to be modeled beyond the simulation resolution. To achieve this, we apply the post-processing models of \citet{Kelley_2017a, Kelley_2017b,sayeb_massive_2021}. These models use inward extrapolation of
host galaxy density profiles and compute the hardening rates of binary inspiral due to dynamical friction, stellar scattering, circumbinary gas-disk driven hardening, and GW emission. \citet{sayeb_mbh_2023} (hereafter \SB{}) explore potential triple MBH systems
by investigating MBHs that experience more than one merger and identifying the subset that are most likely to experience strong triple interactions. We take this sub-population of likely strong triples and model their outcomes by incorporating the triple interaction subgrid model from the \citet{bonetti_post-newtonian_2016,bonetti_post-newtonian_2018} numerical simulations of MBH triplets to model their outcomes.

Throughout this paper, we use the following binary and triple parameters: $m_1$ and $m_2$ denote the masses of the primary and secondary members of the inner MBH binary. The mass ratio of the inner binary is $q_{\rm in} = m_2/m_1$ and is defined to be always smaller than unity. The intruder MBH mass is denoted by $m_3$, and the outer binary mass ratio is $q_{\rm out} = m_3/(m_1 + m_2)$. Note that \qout{} can be greater than or less than unity.  

\subsection{Illustris simulation}

The Illustris simulation suite is a set of cosmological, hydrodynamical simulations run using the {\tt AREPO}  code \citep{Springel_2010}. This code combines the advantages of smooth particle hydrodynamics (SPH) (e.g., \citet{Gingold1977, Lucy1997}) and an Eulerian-mesh-based approach (e.g., \citet{BERGER198964}). The highest resolution run, `Illustris-1' (referred to in this work simply as Illustris), is a cosmological box of side $L_{\text{box}} = 75 h^{-1}$ Mpc. This simulation has a DM resolution of $6.26 \times 10^6 M_{\odot}$ and a baryonic mass resolution of $1.26 \times 10^6 M_{\odot}$ and runs from $z=127$ to $z=0$. The simulations assume a WMAP9 cosmology \citep{Hinshaw_2013} with parameters $\Omega_m = 0.2865$, $\Omega_{\Lambda} = 0.7135, \sigma_{8} = 0.820$, and $H_{0} = 70.4$ \kms Mpc$^{-1}$.

In Illustris, MBH particles are seeded with an initial mass of $1.4 \times 10^5 M_{\odot}$ in halos of mass above $7.1 \times 10^{10} M_{\odot}$ that do not already contain an MBH \citep{Sijacki_2015}. The MBHs grow through Eddington-limited Bondi accretion and mergers \citep{Springel_2005, Di_Matteo_2005}. For an overview of the various sub-resolution physical models, including accretion, feedback, star-formation, chemical enrichment and cooling, see \citet{Vogelsberger_2013} and \citet{Torrey2014}. Galaxies are identified by the Subfind algorithm \cite{Dolag2009}, which locates gravitationally bound `subhalos' within friends-of-friends halos. In this work, we hereafter use the term `MBH host galaxies' to refer to the Subfind subhalos that host MBHs. 

Similar to \SB{}, we use Illustris over the more recent IllustrisTNG \citep{Nelson_2017,Springel_2017,Marinacci2018,Pillepich2018b,Naiman2018,Nelson2019,Pillepich2019}, mainly for continuity with previous work using these MBH inspiral models \citep{Kelley_2017a,sayeb_massive_2021}. It should be noted that the MBH seed mass in IllustrisTNG is greater than Illustris by a factor of $\approx 8$. The fiducial IllustrisTNG box volume is (100 Mpc)$^3$, but it also features both higher-resolution (50 Mpc)$^3$ and lower-resolution ($300$ Mpc)$^3$ volumes. We plan to extend our analysis to additional simulations in future work.

\subsection{Description of the inspiral model}

\label{sec: insp hardening}

The time when the MBH merges in Illustris is called the binary \emph{formation time}, after which we apply the post-processing inspiral models of  \citet{Kelley_2017a,Kelley_2017b}. The MBH population considered here is the same as the one in \SB{} and \citet{sayeb_massive_2021}. Illustris uses a repositioning scheme which pins MBHs to the potential minimum of their host halos. The effects of this are most problematic for small satellite halos that are spuriously re-seeded with MBHs after losing their initial MBH during an interaction with a larger halo. As this predominantly affects MBHs close to the seed mass, we follow previous work and exclude
MBHs below $10^6$ \msun{}. The inspiral model also requires a binary to have a reasonably well resolved host galaxy in the snapshots before and after the merger for calculating the environmental effects for calculating hardening rates. This adds additional constraints of a minimum of 80 gas cells, 80 star particles, and 300 DM particles. All of these constraints reduce the subset of 23708 MBH mergers in Illustris to 9234 systems \citep{Kelley_2017a,sayeb_massive_2021,sayeb_mbh_2023}. The binary inspiral model \citep{Kelley_2017a, Kelley_2017b,sayeb_massive_2021} evolves these binaries through four different hardening mechanisms \citep{Begelman1980}: dynamical friction (DF), loss cone scattering (LC), hardening due to circumbinary disk (CD), and hardening via gravitational wave radiation (GW). 

Our fiducial model assumes a full loss cone and allows for the eccentricity to vary in the LC phase. The LC hardening rate $da/dt$ and the eccentricity evolution $de/dt$ is based on the scattering experiments by \citet{sesana_interaction_2006} and are as follows:

\begin{align}
    \left( \frac{da}{dt} \right)_{\rm LC} &= - \frac{G \rho}{\sigma} a^2 H\\
    \left( \frac{de}{dt} \right)_{\rm LC} &= \frac{G \rho}{\sigma} a H K
    \label{dadt_lc}
\end{align}

where $a$ is the binary separation, $e$ is the orbital eccentricity. $\rho$ and $\sigma$ corresponds to the stellar density and the stellar velocity dispersion profiles of the host in Illustris. $H$ and $K$ are dimensionless constants set by numerical scattering experiments and here we use the values from \citet{sesana_interaction_2006}. The initial eccentricity of all binaries is set to be $e_{0} = 0.6$ at the beginning of the DF phase in our fiducial model. The eccentricity is evolved during the LC and GW-dominated phase. The GW hardening rate and the eccentricity evolution follows \citet{PetersandMathews1964}. Note that, we do not include eccentricity evolution in the CD phase. 

\begin{figure*}[htb!]
    \centering \includegraphics[scale=0.6]{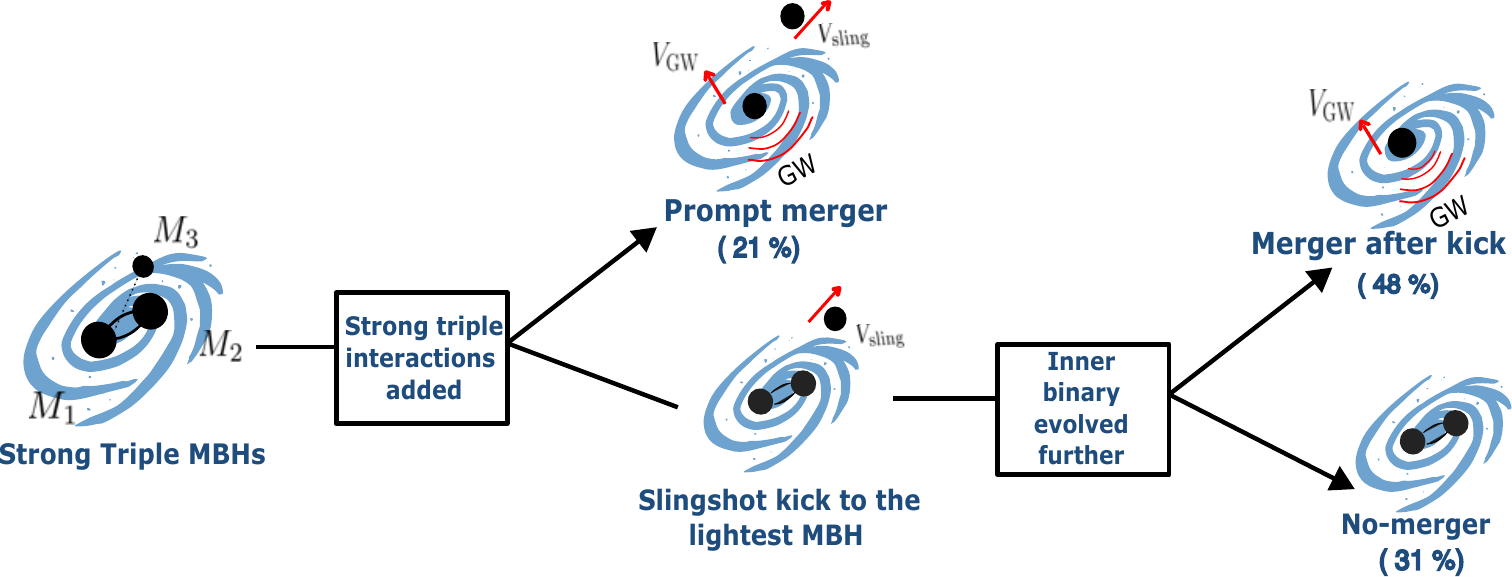}
    \caption{Schematic diagram illustrating the outcomes of binary and triple interactions in the subpopulation of Illustris MBH binaries that form strong triples. Triples interactions are added for the strong triple subpopulation ($a_{\rm triple} < 100$  pc). Merger fractions averaged over 100 realizations are indicated for the three outcomes: i) \emph{prompt merger} (21 \%) ii) \emph{merger after kick} (48 \%) and iii) \emph{no-merger} (31 \%).}
    \label{fig:Triple-outcomes}
\end{figure*}

\subsection{Outcomes of triple interactions}
\label{sec: triple outcomes}

Although the binary inspiral model evolves each binary in isolation, we can identify binary systems sharing a common MBH in the merger tree. If the binary inspiral times of the two binaries overlap, they are considered co-existing binaries and may represent a potential triple system. 

Out of the sample of coexisting binaries, we denote the binary with the earliest formation time as the ``first''or ``inner" binary and the ``second" or ``outer" binary is the one with the later formation time. After comparing the evolution of binary separation vs time $(a,t)$, triples are defined to be coexisting binaries where the first binary is overtaken by the second before $z=0$. For such systems, triple formation time $t_{\rm triple}$ is defined as the point in time when the separation of the outer binary becomes smaller than that of the inner binary, and the corresponding separation is called the triple formation radius $a_{\text{triple}}$. Triples with $a_{\text{triple}} > 100$ pc are classified as ``weak triples'' and those with $a_{\text{triple}} > 100$ pc are ``strong triples''. The cutoff between strong and weak triples is set at 100 pc and is validated through K-means clustering applied to all the triples (\SB{}). Because the strong triples are more likely to undergo three-body interactions, we choose to focus only on strong triples for our exploration of the outcomes of triple interactions. 

Once the triples have been identified from the Illustris binaries, the next step is to find the masses of the BHs in the binary $m_1$, $m_2$ and the intruder BH mass $m_3$. To identify these BHs from the first and second binary, we first calculate the expected mass of the merger product. This is done by summing the masses of the first binary components, $m_1$ and $m_2$, and the total mass accreted by the binary between its formation and the time of triple formation. Let $\Delta m$ represent this accreted mass, calculated using the black hole accretion rates from Illustris. The expected mass of the merger product is then $m_{\rm prod} = m_1 + m_2 + \Delta m$. Among the two BHs in the second binary, we identify the one with mass closest to $m_{\rm prod}$ as the merger product. The other BH is then classified as the intruder, with mass denoted by $m_3$.

We also account for mass growth for the individual BHs in the first binary. To calculate the mass accreted by each BHs in the first binary, we assume that the accretion onto each BH is proportional to their masses, similar to the \texttt{Mod\_Prop} case in \cite{Siwek_2020}. The new primary and secondary masses become:
\begin{eqnarray}
    \label{eq: m1m2 accreted mass}
    m_{1} \rightarrow m_{1} +  \Delta m/(1+q)\\
    m_{2} \rightarrow  m_{2} +  \Delta m  \, q / (1+q)
\end{eqnarray}

After finding the mass growth of each BH of the inner binary, we consider the maximum of the two masses to be the primary and the minimum to be secondary (i.e $m_1 = \max (m_1,m_2)$ and vice-versa). Once we obtain the set of strong triples and their masses ($m_1,m_2,m_3$), we incorporate triple dynamics with the help of results from the triple MBH simulations from \citet{bonetti_post-newtonian_2018-1}. 

The outcomes of triple MBH interactions in \citet{bonetti_post-newtonian_2018-1} are sampled for different values of primary mass $m_1$ sampled in the range [ $10^5$ \msun{}, $10^{10}$ \msun{}] and inner mass ratio $q_{\rm in} \in [0.03,1]$ and outer mass ratio $q_{\rm out} \in [0.03,1]$ . They also perform a set of simulations for the case where the intruder is more massive than the MBHB, i.e., $q_{\rm out} > 1$ systems. The simulation is performed for systems with different inner and outer orbit eccentricities and relative inclinations. For each simulation grid point, three merger fractions for the merger of $m_1-m_2$, $m_1-m_3$, and $m_2-m_3$ represented by $a,b,c$ are recorded. Similar to \citet{bonetti_post-newtonian_2018}, we perform grid interpolation between our set of strong triples represented by ($m_1$, \qin, \qout) and the surveyed points from the simulation to obtain the merger fractions $a,b,c$  for the strong triples. To decide the outcome of the strong triple interaction, a random number P is chosen between 0 and 1, and one of the two choices is considered:

\begin{itemize}
    \item A \emph{prompt merger} of one of the pairs in the triple system triggered by the triple interaction and a the lightest BH kicked out when $P \leq a+b+c$. 

    \item For $P > a+b+c$, the lightest MBH is kicked due to gravitational slingshot from the triple interactions. The remaining binary will evolve further after the kick.
\end{itemize}

For our fiducial model, we perform 100 realizations of the outcomes for the strong triple population. To further evaluate the merger time, we need to evolve the binaries after the triple interaction. For \emph{prompt mergers}, \citet{bonetti_post-newtonian_2018-1} shows that the time spent by strong triples before merger can be fitted with a log-normal distribution, with a mean $\mu = 8.4$ (corresponding to $\simeq 250$ Myr) and standard deviation $\sigma = 0.4$ in $\log(T/yr)$. We add a merger time sampled from this distribution to $t_{\rm triple}$ to get the prompt merger time.

In cases where the lightest BH is kicked from the triple system and there is no prompt merger, the leftover binary may coalesce within a Hubble time, though this is not always the case. \citet{bonetti_post-newtonian_2018-1} saw that the remaining binaries that merge via GW emission constitute  20\% of the total merger fraction in triple systems. In cases of a light intruder, after it gets scattered via gravitational slingshot, the remaining binary is hardened. The new binary separation of the inner binary is then calculated following \citet{volonteri_assembly_2003}. In the case of a heavier intruder that causes an exchange interaction, it would replace the secondary MBH in the inner binary. We therefore evolve the outer binary in this case with the new calculated binary separation. If $t_{\rm merger}<t_{\rm H}$, we call such cases \emph{merger after kick}. The systems that do not merge within Hubble time are stalled and hence added to the \emph{no-merger} case. These outcomes are shown in Figure \ref{fig:Triple-outcomes}. Note that because weak triples (formed at separations $>100$ pc) are unlikely to undergo three-body interactions, we do not include these systems in the calculations of the triple interaction outcomes. However, any mergers between BHs in these systems that occur via normal binary inspiral are counted in the total merger population.

Note that the binary can have a higher orbital eccentricity after the triple interaction. Although eccentricity is included in the binary evolution model, we are not currently evolving them after the triple interaction. In \citet{bonetti_post-newtonian_2018}, eccentricity was assigned to the triples from the eccentricity distribution from the triple MBH recorded at an arbitrary point close to their merger. In future work, we will employ a more accurate eccentricity and inspiral evolution coupled with strong triple interactions.

We also consider a ``stalled" model, where all potential triple systems from Illustris are assumed to stall at the time of binary formation. We then apply the triple interaction subgrid model to these systems, averaging over 100 realizations of the outcomes and look for prompt mergers. This model, similar to the \emph{Model-stalled} in \citet{bonetti_post-newtonian_2018}, allows us to assess the role of triple interactions in facilitating mergers when binaries would otherwise remain stalled.

\subsection{Gravitational wave recoil kick calculation}
\label{sec: GW recoil kick}

The merger of two MBHs with unequal masses or spins will result in a GW recoil to the merger remnant. The GW recoil velocity depends on the spins and the mass ratio of the merging BHs.  The mass ratio is calculated for the merging BHs and the BH spins are drawn from physically motivated distributions. If the inspiral is driven by torques from a circumbinary gas disk, the BH spins tend to align with the orbital angular momentum preferentially \citep{Bogdanovic_2007, Dotti2010,Miller_2013}. Whereas, in gas poor systems where the inspiral is primarily driven by stellar interactions, the resulting spin orientations may tend to be random \citep{Merritt_2012}. As we are unsure of the efficiency of these processes in various merger environments, we pick three spin distributions for different spin orientations and magnitudes for the MBH population. 

\begin{figure}[htb!]
    \centering
    \includegraphics[scale=0.42]{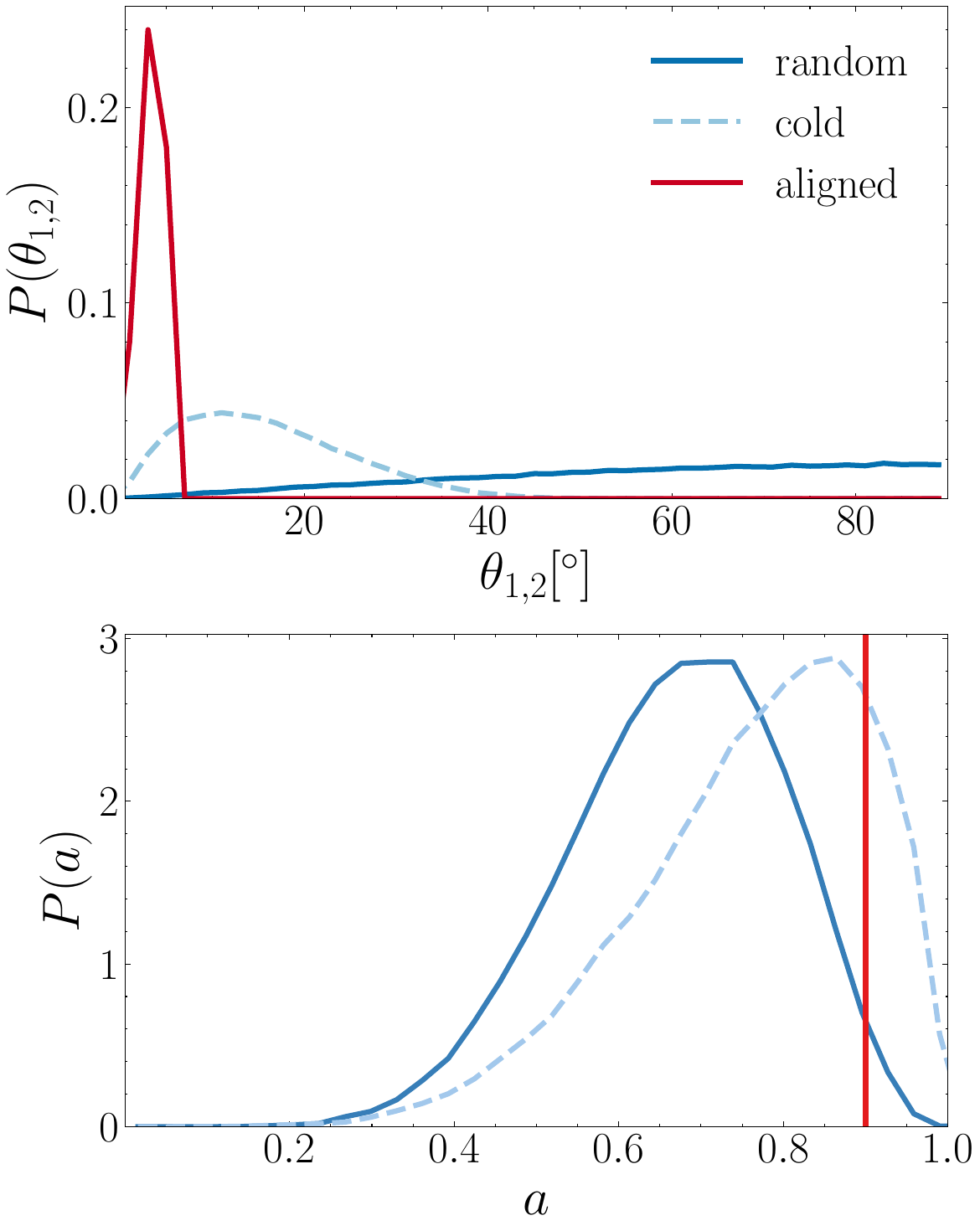}
    \caption{Normalized distributions of the three models for pre-merger MBH spins considered in this work: \texttt{random} (solid dark blue), \texttt{cold} (dotted light blue), and \texttt{aligned} (solid red). \textit{Top panel:} misalignment angle distribution of the spins. \textit{Bottom panel:} spin magnitude distribution of the models.  
    }
    \label{fig:spin-dist}
\end{figure}

The three spin models considered here are from \cite{Blecha2016}. The first spin model produces a random spin distribution as a result of successive dry mergers. The spin magnitude is represented by a beta distribution that peaks at $a \sim 0.7$ \citep{Lousto_2012} and has a large tail towards low spins. The spin misalignment angle $\theta$ (relative to the orbital angular momentum) is drawn randomly in the interval $(0^{\circ},180^{\circ})$. This model is called \texttt{random-dry} but for the sake of simplicity we will refer to it as \texttt{random}. It is represented by the dark-blue curve in Figure \ref{fig:spin-dist}. At the other extreme, we can assume the spins are always highly aligned before the merger. To represent this choice, we consider the misalignment angle chosen over the interval $(0^{\circ},5^{\circ})$ and a high spin magnitude ($a = 0.9$). This spin model is called the \texttt{aligned} and is represented by the red lines in Figure \ref{fig:spin-dist}.

For the third spin model, we take the cold gas fraction of the progenitor halos into consideration. As BH spins may be aligned due to the presence of a circumbinary gas disk, we assume that gas-rich galaxy mergers likely trigger the formation of a massive circumbinary disk. We define a ``gas-rich" galaxy as one with cold (star-forming) gas fraction $f_{\text{gas,sf}}> 10 \%$. The gas fraction is defined as $f_{\text{gas,sf}} = M_{\text{gas,sf}}/(M_{*} + M_{\text{gas,sf}})$ where the $M_{\text{gas,sf}}$ and $M_{*}$ are the gas mass and stellar mass within the stellar half-mass radius of the progenitors. The majority of galaxies in our population (more than $85 \%$) are gas-rich. The gas-rich mergers would then have BH spins drawn from a \texttt{cold} spin evolution model described in \cite{Dotti2010} and \cite{Lousto_2012}. This model represents the case of the nearly aligned spin ($\lesssim10^{\circ}$) and is represented by the dotted blue lines in figure \ref{fig:spin-dist}. For gas-poor mergers with $f_{\text{gas,sf}} <  10 \%$, the spins are drawn from the \texttt{random} described above. We call this the \texttt{hybrid} spin model.

Once the progenitor BH spins are drawn from the distribution for a given spin model, the recoil kick velocity is calculated using a fitting formula from \cite{Lousto_2012}, which is based on numerical relativity simulations:

\begin{equation}
{\mathbf v_{\rm recoil}} = v_{\rm m}  {\mathbf {\hat e}_{\perp,1}}+ v_{\perp} ({\rm cos} \,\xi \, {\mathbf {\hat e}_{\perp,1}} + {\rm sin} \,\xi \, {\mathbf {\hat e}_{\perp,2}}) + v_{\parallel} {\mathbf {\hat e}_{\parallel}},
\label{eqn:kick}
\end{equation}
\begin{equation}
v_{\rm m} = A \eta^2 \sqrt{1 - 4\eta} \,(1 + B \eta), 
\end{equation}
\begin{equation}
v_{\perp} = \frac{H \eta^2 }{(1 + q )}(a_{2\parallel} - q a_{1\parallel}),
\end{equation}
\begin{eqnarray}
v_{\parallel} = \frac{16 \eta^2}{(1+ q)} \left [ V_{1,1} + V_{\rm A} \tilde S_{\parallel} + V_{\rm B} \tilde S_{\parallel}^2 + V_{\rm C}\tilde S_{\parallel}^3 \right ] \times \\ \nonumber 
|\, {\mathbf a_{2\perp}} - q {\mathbf a_{1\perp}} | \, {\rm cos}(\phi_{\Delta} - \phi_1),
\end{eqnarray}

where $\eta \equiv q/(1+q)^2$ is the symmetric mass ratio, $\perp$ and $\parallel$ refer to vector components perpendicular and parallel to the orbital angular momentum, respectively, and ${\mathbf {\hat e}_{\perp,1}}$ and ${\mathbf {\hat e}_{\perp,2}}$ are orthogonal unit vectors in the orbital plane. The vector ${\mathbf {\tilde S}} \equiv 2({\mathbf a_2} + q^2 {\mathbf a_1})/(1+q)^2$, and $\phi_{\Delta}$ is the angle between the in-plane component ${\mathbf \Delta_{\perp}}$ of the vector ${\mathbf \Delta} \equiv M^2({\mathbf a_2} - q {\mathbf a_1})/(1+q)$ and the infall direction at merger. The phase angle $\phi_1$ depends on the initial conditions of the binary and is assumed to be random. The best-fit values of $A = 1.2\times 10^4$ \kms, $B = -0.93$, $H = 6.9\times10^3$ \kms, and $\xi = 145^{\circ}$ are taken from \cite{Gonz_lez_2007} and \cite{Lousto2008}, and the coefficients $V_{1,1} = 3677.76$, $V_{\rm A} = 2481.21$, $V_{\rm B} = 1792.45$, and $V_{\rm C} = 1506.52$ (all in km s$^{-1}$) are defined in \cite{Lousto_2012}.

\subsection{Slingshot kick calculation}
\label{sec: triple kick calculation}

If the intruder BH mass ($m_3$) is smaller than both the BHs ($m_1,m_2$) in the inner binary, then the triple encounter will likely lead to the scatter of the intruder BH. The intruder is kicked due to gravitational slingshot and the binary recoils by momentum conservation. The binary orbit shrinks as a result of the encounter. The binding energy for such encounters increases by the amount $\langle\Delta E/ E_B\rangle = 0.4 (m_3/(m_1 + m_2))$ \citep{Hills1980}, where $E_B$ is the binding energy of the binary. 

In cases where the intruder MBH is more massive than the binary components ($m_3 > m_2$ or $m_3 > m_1$), there will likely be an exchange event with the lightest BH ejected and the remaining two forming a binary. In both the cases there is an increase in the binding energy of the BH pair. For outer mass ratios \qout $\leq 2$, most of the increase in binding energy is due to shrinking of the orbit whereas for \qout $> 2$, the increase is mainly due to the change in mass due to the exchange of a low mass member by a more massive BH \citep{volonteri_assembly_2003}. 

The binding energy $E_B$ is calculated at the triple formation time when the strong interactions take over. The separation at the triple formation time, $a_{\rm triple}$ would be an overestimate for the actual separation where ejection happens. The binary could instead evolve to the hardening radius $a_h$ \citep{quinlan_dynamical_1996,Begelman1980}. To account for this, we take the minimum of the two separations for the separation to get the separation during gravitational slingshot recoil, $a_{0} = \min(a_h, a_{t})$.

The change in binding energy is calculated by the same scheme used by \cite{volonteri_assembly_2003}. The scheme is summarized as follows:

\begin{itemize}
    \item If $m_3<m_2$, $m_3$ is scattered of leaving with $\langle\Delta E/ E_B\rangle = 0.4 \, q_{\rm out}$  and a new semi-major axis $a_1 = a_0/(1+0.4 \, q_{\rm out})$
    \item If $m_3 > m_2$ and \qout $< 2$, it will result with an exchange event with $\langle\Delta E/ E_B\rangle = 0.4 \, q_{\rm out}$ 
    and a new semi-major axis $a_1 = a_0 (m_3/m_2)/(1+0.4 \, q_{\rm out})$
    \item If $q_{\rm out} > 2$, an exchange happens with $\langle\Delta E/ E_B\rangle = 0.9$ and a new semi-major axis $a_1 = 0.53 a_0 (m_3/m_2)$.
\end{itemize}

The kinetic energy of the ejected BH through gravitational slingshot will then be:

\begin{equation}
    K_{\rm ej} = \frac{\Delta E}{1 + (m_{\rm ej}/m_{\rm bin})}
    \label{eq: K_ej}
\end{equation}

where $m_{\text{ej}}$ is the mass of the lightest BH among the three and $m_{\text{bin}} = m_1 + m_2$ is the mass of the final binary. Hence, during scattering $m_{\text{ej}} = m_{3}$ and $m_{\text{bin}} = m_1 + m_2$ and during exchange  $m_{\text{ej}} = m_{2}$ and $m_{\text{bin}} = m_3 + m_1$. The gravitational slingshot kick velocity is calculated from equation \ref{eq: K_ej} and the orbital separation after the strong triple interaction is updated. The binary after the kick is evolved further if it didn't merge via through prompt-merger (as described in Section \ref{sec: triple outcomes}). If a binary remains after the triple interaction, this updated separation is used as the starting point to model its continued inspiral.

\subsection{MBH ejections}
\label{sec: BH ejections}

The kicked BH needs to have a velocity larger than the escape velocity $v_{esc}$ of the host galaxy for it to be ejected. To count the number of possible ejections from both GW kicks and slingshot kicks, the escape velocity of the host is calculated from its potential. For the DM component, we assume a simple \cite{Hernquist1990} potential $\Phi(r) = - G M_{DM} / (r + r_H)$, where $M_{DM}$ is the sum of the DM masses of the progenitor subhalos of the merging MBHs and $r_H$ is the scale radius. $r_H$ is related to the half mass radius by $r_{\rm half} \approx 2.414 \, r_{H}$  .The stellar component is modeled using a softened isothermal density profile, $\rho_{*} = \sigma_{*}^{2}/(2 \pi G (r^2 + r^2_{\text{soft}}))$, with velocity dispersion $\sigma^2_{*} = G M_{*}/R_{\text{bulge}}$ and $r_{\text{soft}} = G M_{BH}/\sigma^{2}_{*}$. We take the stellar bulge truncated at an outer radius $R_{\text{bulge}}$ to be a simple average of the progenitor stellar half-mass radii. The potential of merging galaxies is however dynamic and major mergers in particular can drastically change the structure of the progenitors. \cite{blecha_recoiling_2011} have shown that the escape velocity \vesc is volatile during a merger and in some cases \vesc shows a large increase, during the merger.

A BH is counted as ejected if the recoil kick from either GW recoil or slingshot kick is larger than the central escape speed of the host galaxy. An ejected BH's ID is checked for repetition in the merger tree and any subsequent merger that this BH undergoes is discarded.

\section{Results}

\label{sec: results}

\subsection{Properties of the merger population}

\begin{figure*}[!htb] 
    \centering
    \includegraphics[scale=0.52]{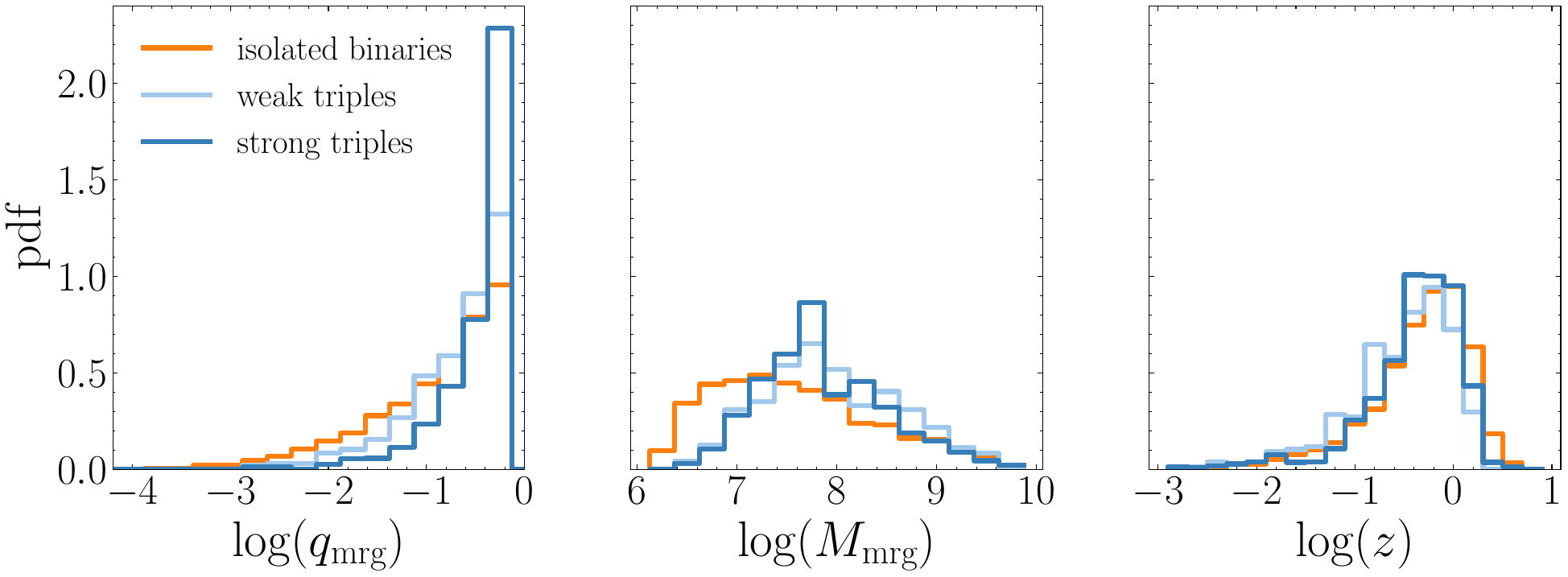}
    \caption{Distributions of the merging BH mass ratio ($q_{\text{mrg}}$), total mass ($M_{\text{mrg}}$), and redshift ($z$) of the merging MBHs in the isolated binaries, weak triples, and strong triples subpopulations. The median mass ratio of mergers in strong triples, weak triples, and isolated binaries is 0.63, 0.40, and 0.25, respectively. $64 \%$ of strong triple mergers have a $q_{\rm mrg} > 0.1$. The median total mass of mergers in strong triples, weak triples, and isolated binaries is $7.87 \times 10^7$ \msun{}, $9.72 \times 10^7$ \msun{} and $3.89 \times 10^7$ \msun{}, respectively.}
    \label{fig:merger-properties-hist}
\end{figure*}

The fiducial model has 9234 binary systems, out of which 22 \% (2029) form triple MBH systems. The remainder, which we call isolated binaries, are systems that evolve entirely in isolation. Of the 2029 triples in the population,
520 (26 \%) are strong triples and 1509 (74 \%) are weak triples (\SB{}). The mergers in isolated binaries are found by tracking their evolution via the inspiral model described in Section \ref{sec: insp hardening}. Mergers induced by strong triple interactions are identified for 100 realizations of each strong triple system, as described in Section \ref{sec: triple outcomes}. The mergers in weak triples are determined by the inspiral timescale of the inner binary. We do not account for strong triple interactions in the weak triple subpopulation.

In Figure \ref{fig:merger-properties-hist}, we show the normalized mass ratio ($q_{\text{mrg}}$), the total mass ($M_{\text{mrg}}$), and the redshift ($\log z$) of the merging MBHs in the three distinct subpopulations. The median mass ratio for mergers resulting from strong triples, $q_{\text{mrg}} = 0.63$, is significantly higher than the mass ratio of isolated binary mergers ($q_{\text{mrg}} = 0.25$). About 64 \% of strong triple mergers are major mergers with $q_{\text{mrg}} > 0.1$. Mergers resulting from triple interactions are also more massive than isolated mergers; the median mass of merging MBHs in isolated binaries is $3.89 \times 10^7 M_{\odot}$ versus $7.87 \times 10^7 M_{\odot}$ for strong triples.
This is largely because massive galaxies (hosting more massive MBHs) experience more mergers and are thus more likely to host triple MBH systems.
For the weak triples, the median mass ratio is 0.40, and the median  $M_{\text{mrg}} = 9.72 \times 10^7 M_{\odot}$. This reflects the finding in \SB{}  that among triples, weak triples are found in massive systems and have lower mass ratios compared to strong triples. Such systems often occur when two satellite galaxies (each carrying a low-mass MBH) fall into the same massive central galaxy.

21 \% of strong triple interactions result in a \emph{prompt merger} of two BHs and a slingshot kick to the other BH, which is often the lightest in the system.  In other cases, a slingshot kick to the lightest BH may not be accompanied by a prompt merger of the remaining BHs, but this remaining binary could still subsequently merge, often on a shorter timescale as a result of the triple interaction. Such \emph{merger after kick} events occurs in 48 \% of strong triples. Thus, a total 69 \% of strong triple systems result in a merger event by $z=0$. Before including triple dynamics, we find that only 40\% (210) of strong triple systems have an inner binary resulting in a merger via isolated binary inspiral evolution. Thus, strong triple dynamics raise the merger fraction of this binary sub-population from 40\% to 69 \%. Overall, strong triples increase the merger fraction of the total population by 4 \%.

The merger fraction seen in the \emph{merger after kick} scenario is much higher compared to the \cite{bonetti_post-newtonian_2018} simulations. This is because the post-ejection mergers in their simulations are calculated by considering only GW-induced inspiral. We evolve the binary using our hardening model initialized from an updated, closer binary separation after the triple interaction, as described in Section \ref{sec: triple outcomes}. This leads to many cases where the leftover binary coalesces after the slingshot kick. 

We do not consider the eccentricity evolution of the inner binaries. Triple interactions can make the binary extremely eccentric and in \cite{bonetti_post-newtonian_2019}, they assign eccentricities to the triples from the distribution found through the simulations. This could impact the evolution of the remnant binary following a gravitational slingshot kick. The kicked MBH could also fall back into the galactic potential and merge with the binary merger remnant in some cases \citep{hoffman_dynamics_2007}. This could potentially provide some additional mergers in strong triples.

\begin{figure*}[!htb] 
    \centering
    \includegraphics[scale=0.44]{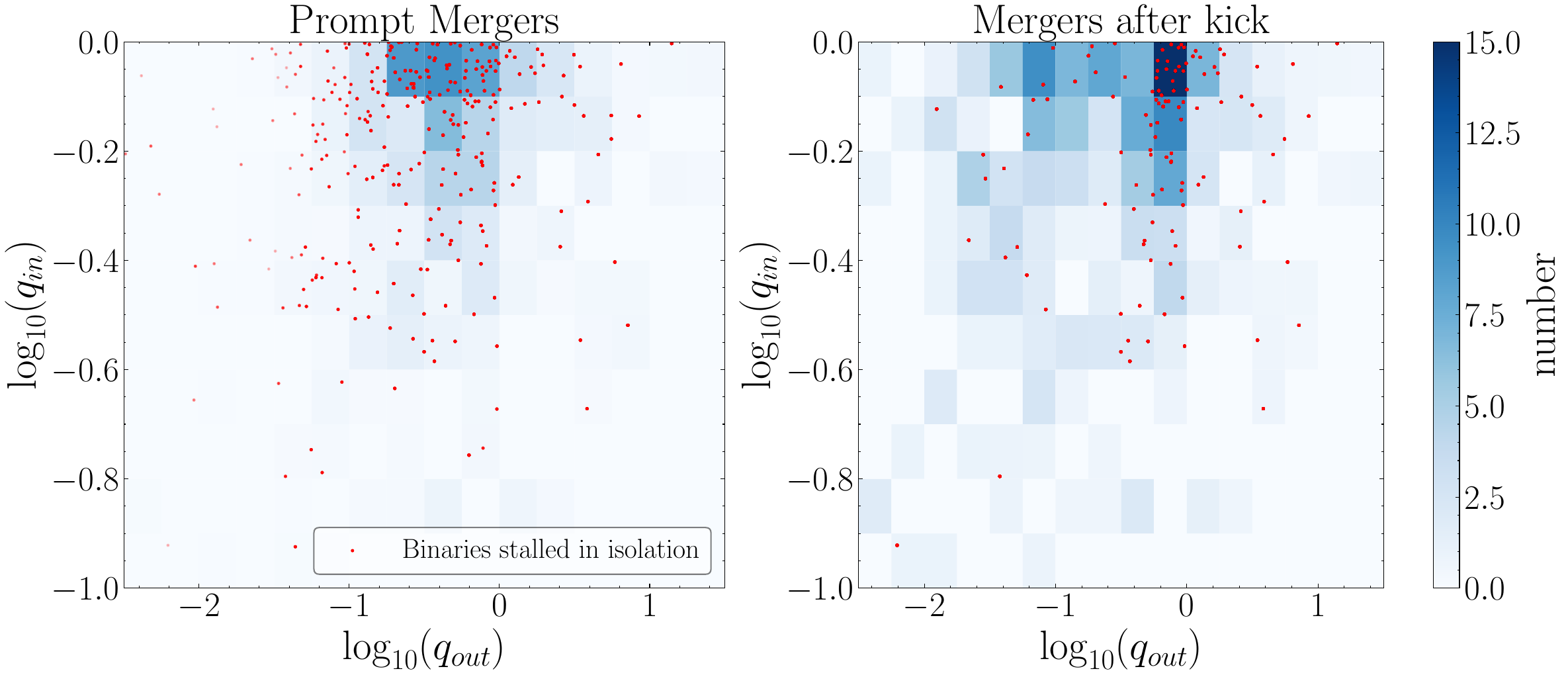}
    \caption{A 2D histogram in the $\log q_{\text{in}} - \log q_{\text{out}}$ plane for 
    prompt mergers (left panel) and 
    merger after a slingshot kick (right panel) from triple interactions. The red dots correspond to triple-induced mergers
    where the inner binary would {\em not} have merged in isolation, and they are plotted for all realizations of the strong triple outcomes. The distribution highlights that the prompt mergers is predominantly seen in equal mass triples. Additionally, prompt mergers involve a higher fraction of cases where the binaries would have otherwise stalled in isolation compared to mergers kick.}
    \label{fig:2d-hist-qin-qout}
\end{figure*}

Figure \ref{fig:2d-hist-qin-qout} illustrates the number of mergers in the $\log q_{\rm in}$ - $\log q_{\rm out}$ plane for the \emph{prompt merger} and \emph{merger after kick} cases in strong triples. The prompt mergers peak around equal-mass triplets ($\log q_{\rm in} \sim 0$ and $\log q_{\rm out} \sim -0.3) $, similar to what is observed in \citet{bonetti_post-newtonian_2018}. This is because comparable-mass triples will have the strongest interactions, as opposed to systems with less massive intruders that may be ejected with minimal perturbation to the inner binary. We also show the distribution of triple systems that would not have merged in isolation in the same plot (red dots). Note that \qout $ \, > 1$ always corresponds to an exchange event where the intruder replaces the secondary in the inner binary. We see that $60 \%$ of prompt mergers involve binaries that would have otherwise stalled in isolation, compared to $34 \%$ for mergers after ejection, that would have stalled under isolated conditions.

\subsection{MBH Merger rates}

\label{sec: merger rates}

\begin{figure*}[!htb] 
    \centering
    \includegraphics[scale=0.42]{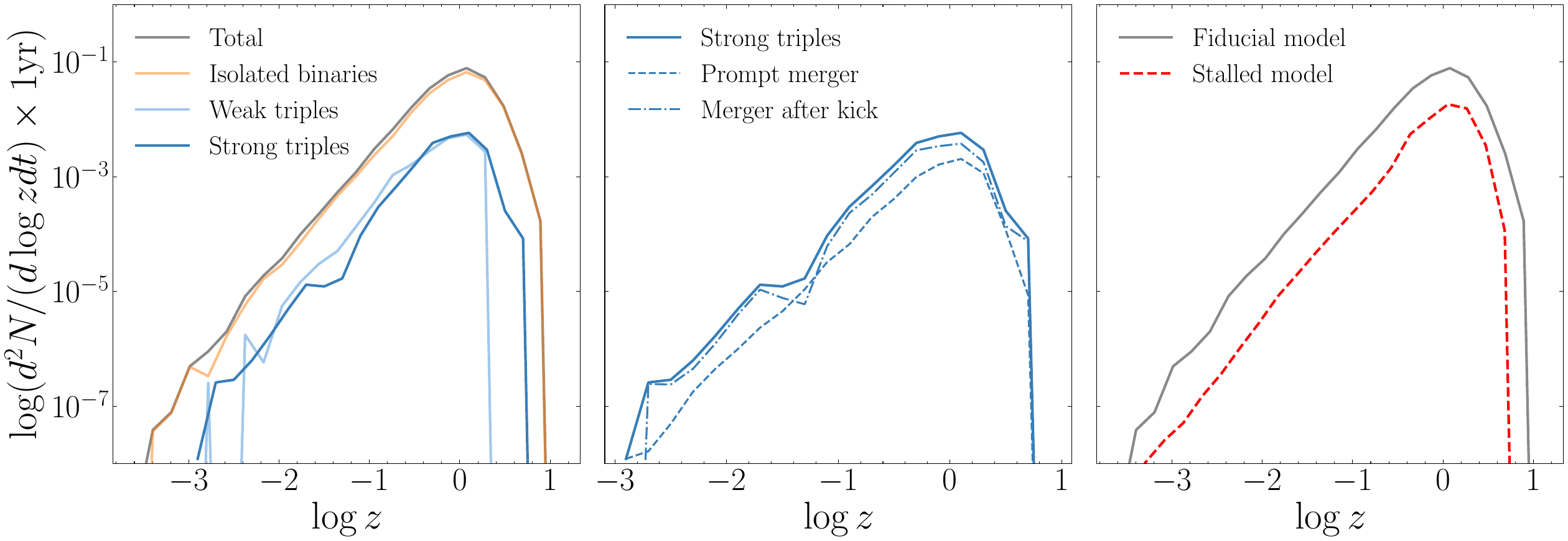}
    \caption{Merger rates of versus redshift for various subpopulations of MBHBs. The left panel shows the total merger rate (gray line) as well as the merger rate for the subpopulations of isolated binaries (orange), weak triple systems (light blue), and strong triples (dark blue). The middle panel shows the merger rate for strong triples, divided into those resulting from the \emph{prompt merger} versus the \emph{merger after kick} scenarios. In the right panel, the merger rate for the stalled model (red dashed line) is shown in comparison to the fiducial merger rate (gray).
    The cumulative merger rates observed at $z=0$ for isolated binaries, weak triples and strong triples are $0.37 \, \rm yr^{-1}$, $0.017 \, \rm yr^{-1}$ and $0.023 \, \rm yr^{-1}$, respectively. The merger rate in the stalled model is suppressed by a factor of $\sim 5$ compared to the fiducial model.}     
    \label{fig:merger-rates}
\end{figure*}

We now focus on the merger rate of MBHs in the total population and subpopulations and assess the impact of the triple interactions on the overall merger rate. We calculate the differential merger rate as follows: 

\begin{equation}
\frac{d^2N}{d \log{z} \, dt_{\text{obs}}} = \frac{1}{1+z} \frac{dN}{d \log{z}} \left( \frac{1}{V_c}\frac{d V_c}{dz}\right) \frac{d z}{dt}    
\label{eq: merger rate}
\end{equation}

where $d^2N/(d \log{z} \, dt_{\text{obs}})$ is the number of merging binaries that would be observed between $\log{z}$ and $\log{z} + d \log{z}$, over a time-interval $dt_{\text{obs}}$ in the observer frame. $\frac{dN}{d \log{z}} \left( \frac{1}{V_c}\frac{d V_c}{dz}\right)$ is the co-moving number density of mergers in the simulation per unit $\log{z}$ interval. The $\frac{1}{1+z}$ factor converts the time interval from the rest frame of the merger to the observer's frame. For the sake of simplicity, we will refer to this as ``merger rates".

Figure \ref{fig:merger-rates} shows the merger rates for the fiducial and stalled models along with the sub-population of strong triples, weak triples, and isolated binaries. The solid grey line is the fiducial model's total merger rate. In the fiducial model, the isolated binaries dominate the mergers as expected as they are $78 \%$ of the population. The total merger rate of isolated binaries is $0.13$ yr$^{-1}$ and the cumulative merger rate observed at $z=0$, for isolated mergers over all cosmic time, is $0.37$ yr$^{-1}$. The solid dark blue line corresponds to the merger rate of strong triples. The redshifts for the strong triple mergers are found from the time of their merger, which is calculated as described in Section \ref{sec: triple outcomes}. The cumulative merger rate (at $z=0$) in strong triples is  $ 0.023$ yr$^{-1}$. The solid light blue line represents the binary mergers in weak triple systems. They produce a cumulative merger rate of $ 0.017$  yr$^{-1}$. 

It is interesting to note that the total strong triple merger rate is comparable to the binary merger rate in weak triple systems, even though the number of strong triple systems ($520$) is lower than the weak triple systems ($1509$) and thus shows the efficiency of strong triple interactions in producing mergers. The total merger rate with strong triple interaction is $0.15$ yr$^{-1}$ and the cumulative merger rate at $z=0$ over all cosmic time is $0.40$ yr$^{-1}$. The merger rates are in agreement with \citet{Blecha2016,Katz2020,sayeb_massive_2021} and \emph{Model-delayed} in \cite{bonetti_post-newtonian_2019}. 

It should be also noted that some of the isolated binaries were part of coexisting binary systems that didn't form a triple (referred to as `failed triples' in \SB{}). In the majority of these systems, the inner binary merges before the outer binary overtakes it. It is possible that some of these systems would have come close to forming a triple system. Of these, \SB{} looks for systems with a small binary separation ratio ($1 < (a_2/a_1)_{\text{min}} \leq 10$) and also a close inner binary at the point of closest approach ($a_1 < 100$ pc) and such systems are only 1\% of the total binary population. These near-misses therefore at best could slightly enhance
the strong triple merger rates. 

In the stalled model, all 9234 binaries are assumed to stall unless they merge via triple interactions. $42$ \% of all MBHs are involved in more than one merger over a Hubble time. We consider strong triple interactions in all such systems, assuming that the intruder BH comes close enough to form a triple system. This model is similar to the \emph{Model-stalled} in \cite{bonetti_post-newtonian_2019}, where the MBHs are stalled at the hardening radius and only allowed to merge through three-body interactions. We find that a total of 812 systems (9 \% of the total population) undergo a prompt merger triggered via triple interactions. The total merger rate of triple systems in the stalled model is $0.032$ yr$^{-1}$. We compare the merger rates of the stalled model (dotted red line) to the fiducial model (solid grey line) in Fig \ref{fig:merger-rates}. Compared to the total merger rate of the fiducial model, this merger rate is only suppressed by a factor of 4.6. \cite{bonetti_post-newtonian_2019} also estimate similar merger rates for their \emph{Model-stalled} and \emph{Model-delayed} models. They also calculate the GWB background in both of their models at the relevant frequency range for PTAs and see that the GWB in the stalled model is only suppressed by a factor of 2-3. Therefore, we verify that even if all the binaries stall, triple interactions prevent the merger rate from dropping too low to produce an observable GWB.

\subsection{Massive major merger subpopulation}

\begin{figure}[!htb] 
    \centering
    \includegraphics[scale=0.55]{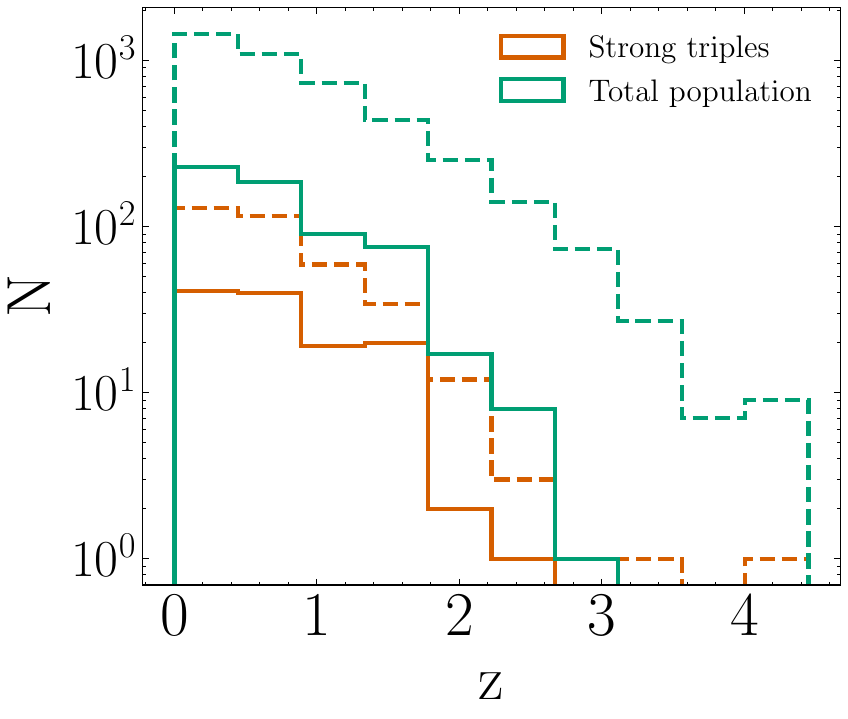}
    \caption{The number of mergers versus redshift is shown for subpopulations of the Illustris MBH merger population. The green dotted line shows the total number of mergers, while the orange dotted line shows the mergers in strong triples. The solid green line show the subset of these mergers that are both massive ($M_{\text{mrg}} > 10^8 M_{\odot}$) and major ($q_{\text{mrg}} > 0.1$). The solid orange line shows mergers in strong triples for the massive major merger subset. Strong triple mergers contribute to 19\% of the massive major merger population.
    }
    \label{fig:massive-major-mergers}
\end{figure}

Although we are not computing the GWB associated with strong triples in this work, we investigate the presence of massive major mergers in the population. Massive, major MBH mergers will be the loudest GW events observable with PTAs (e.g., \citet{sesana_low-frequency_2004}). Here, we look at $M_{\text{mrg}} > 10^8 M_{\odot}$ and $q_{\text{mrg}} > 0.1$ as the massive major subpopulation. 

Strong triples contribute to $19$ \% of the massive major mergers. This is in contrast to the $6 \%$ contribution of the strong triples observed in the total population. In Figure \ref{fig:massive-major-mergers} the number of massive, major mergers for strong triples and the total population are shown in solid lines, while the dotted line indicates all mergers for both populations. The relative contribution of massive, major mergers by strong triples (solid-orange) is seen to be higher than their contribution in total mergers. However, we do not observe a strong redshift dependence in the fractional contribution of strong triple mergers in massive major mergers. 

These results show that massive major mergers are $>3$ times more likely to be driven by strong triple interactions than MBH mergers in general, which suggests that triples may contribute significantly to the MBH merger events observed in PTAs. This result also reflects the distribution of the $M_{\text{mrg}}$ and $q_{\text{mrg}}$ of the triples compared to the isolated binaries in Figure \ref{fig:merger-properties-hist}. Triple interactions are therefore important for massive major and this is because massive galaxies have higher chances of experiencing a second merger in their lifetimes. \citet{bonetti_post-newtonian_2018} showed that triple interactions are more important at higher mass mergers. They saw that half of the mergers with chirp mass $\mathcal{M} > 10^9 $ \msun{} are from the triple channel.

\subsection{Recoil events}

\begin{figure*}[!htb]
    \centering
    \includegraphics[scale=0.6]{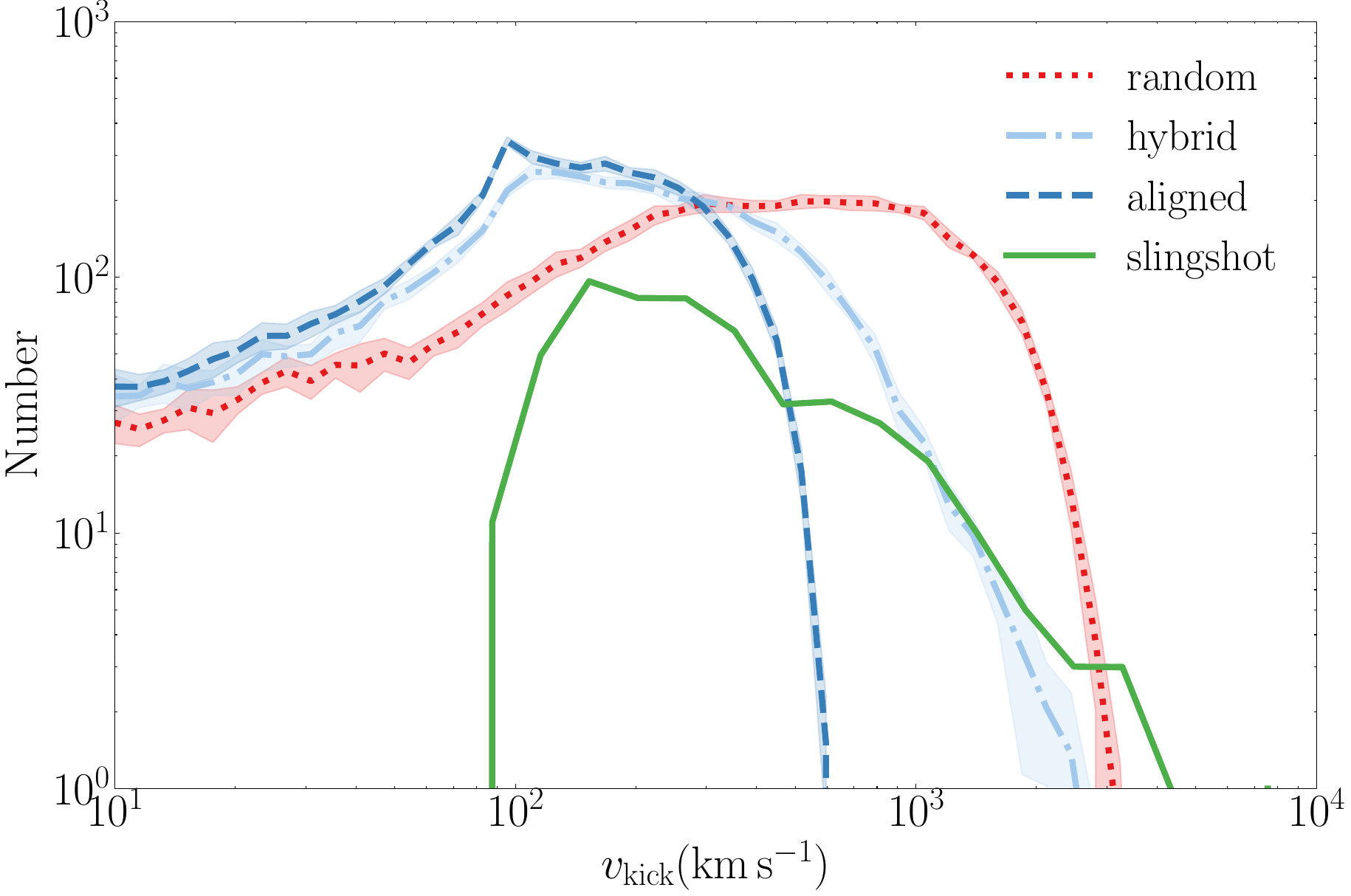}
    \caption{The velocity distributions are shown for GW recoil and slingshot kicks to MBHs in our model. GW recoil velocities are shown for each spin model (\texttt{random}, \texttt{aligned}, \& \texttt{hybrid}) and are averaged over all realizations of pre-merger MBH spins. The mean recoil velocity is plotted with a shaded region indicating the $1\sigma$ spread. Slingshot kick velocities are averaged over all realizations of strong triple interaction outcomes. GW recoil from random spins produces the highest median kick velocity (314 \kms), followed by slingshot kicks (283 \kms). $16\%$ of GW recoil kicks are above 1000 \kms in the random spin model, compared with $10\%$ of slingshot kicks.}
    \label{fig:kick-distributions-for-GW-recoil-and-slingshot}
\end{figure*}

\begin{table*}
\centering
\hspace{-2cm}
\begin{tabular}{|l|l|l|l|l|l|}
\hline
\textbf{Kick type} & \textbf{Median $v_{\text{k}}$} & \textbf{Maximum $v_{\text{k}}$} & \textbf{ $ v > 1000$ \kms} & \textbf{ $v > 500$ \kms} & \textbf{\% ejected} \\ 
 & [km\,s$^{-1}$] & [km\,s$^{-1}$] & [\%] & [\%] & [\%] \\
\hline
GW recoil - \texttt{Random}            & 314                             & 3519                 & $15.6 \pm 0.5$                    & $36.4 \pm 0.6$                     & $15.6 \pm 0.5$                     \\ \hline
GW recoil - \texttt{Hybrid}             & 143                                & 3088                  & $1.2 \pm 0.2$                     & $10.2 \pm 0.3$                      & $1.5 \pm 0.1$                 \\ \hline
GW recoil - \texttt{Aligned}            & 112                                 & 629             & 0                     & $0.64 \pm 0.11$                      & $0.25 \pm 0.06$                      \\ \hline
Slingshot kick         &283                                 & 8368                  & $10.38 \pm 0.03$                    &   $24.1 \pm 0.3$                    & $6.72 \pm 0.04$                      \\ \hline
\end{tabular}

\caption{Statistics of the GW recoil kicks for the three different spin models and slingshot kicks. The values reported are averaged over 10 realizations of the spins.}
\label{table:recoil-statistics} 
\end{table*}

We now turn our attention to the recoil events in our population, happening due to GW recoil and slingshot kicks. As described in Section \ref{sec: GW recoil kick}, we have considered three different spin configurations for our BHs: \texttt{random}, \texttt{hybrid}, and \texttt{aligned} spins. We calculate the GW recoil kick assuming spins drawn from each of these distributions, averaging over 10 realizations per case. For triple systems, the slingshot kick velocity is calculated through the method outlined in Section \ref{sec: triple kick calculation}. We also assign GW recoil kicks for the merged BH in the triple systems by the same procedure as isolated binaries. 

Figure \ref{fig:kick-distributions-for-GW-recoil-and-slingshot} shows the distribution of kick velocities obtained for our merger population with each spin model
- \texttt{random}, \texttt{aligned}, \texttt{hybrid} - as well as the velocity distribution for slingshot kicks. GW recoil kicks are shown as averaged distributions over 10 spin realizations, with a 1-$\sigma$ spread indicated. The maximum kick velocities produced by GW recoil are 3519 \kms\ for \texttt{random} spins, 3088 \kms\ for \texttt{hybrid} spins, and 629 \kms\ for \texttt{aligned} spins, averaged over the realizations. Slingshot kicks yield a maximum kick magnitude of 8368 \kms. Both GW recoil kicks from \texttt{random} spins and slingshot kicks dominate the high-velocity range, with 16\% and 10\% of their kicks exceeding 1000 \kms, respectively. In contrast, only $1$ \% of kicks in the hybrid spin model are above 1000 \kms and none are above this threshold for the \texttt{aligned} spins. The median GW recoil velocities are 314 \kms, 143 \kms, and 112 \kms\ for \texttt{random}, \texttt{hybrid}, and \texttt{aligned} spins, respectively, while slingshot kicks yield a median velocity of 283 \kms.

These results are summarized in Table \ref{table:recoil-statistics}. Although slingshot kicks can reach extreme velocities, their overall occurrence is limited, as strong triples comprise only 6\% of the population. The number of GW recoil kicks from the \texttt{aligned} spin model falls below the number of slingshot kicks at $\sim 600$ \kms. At velocities exceeding 3519 \kms, slingshot kicks surpass those from the GW recoil kicks with \texttt{random} spins. Kicks above this threshold are likely due to gravitational slingshot effects from triple MBH interactions.

\subsection{Ejection events}

\begin{figure*}[!htb]
    \centering
    \includegraphics[scale=0.5]{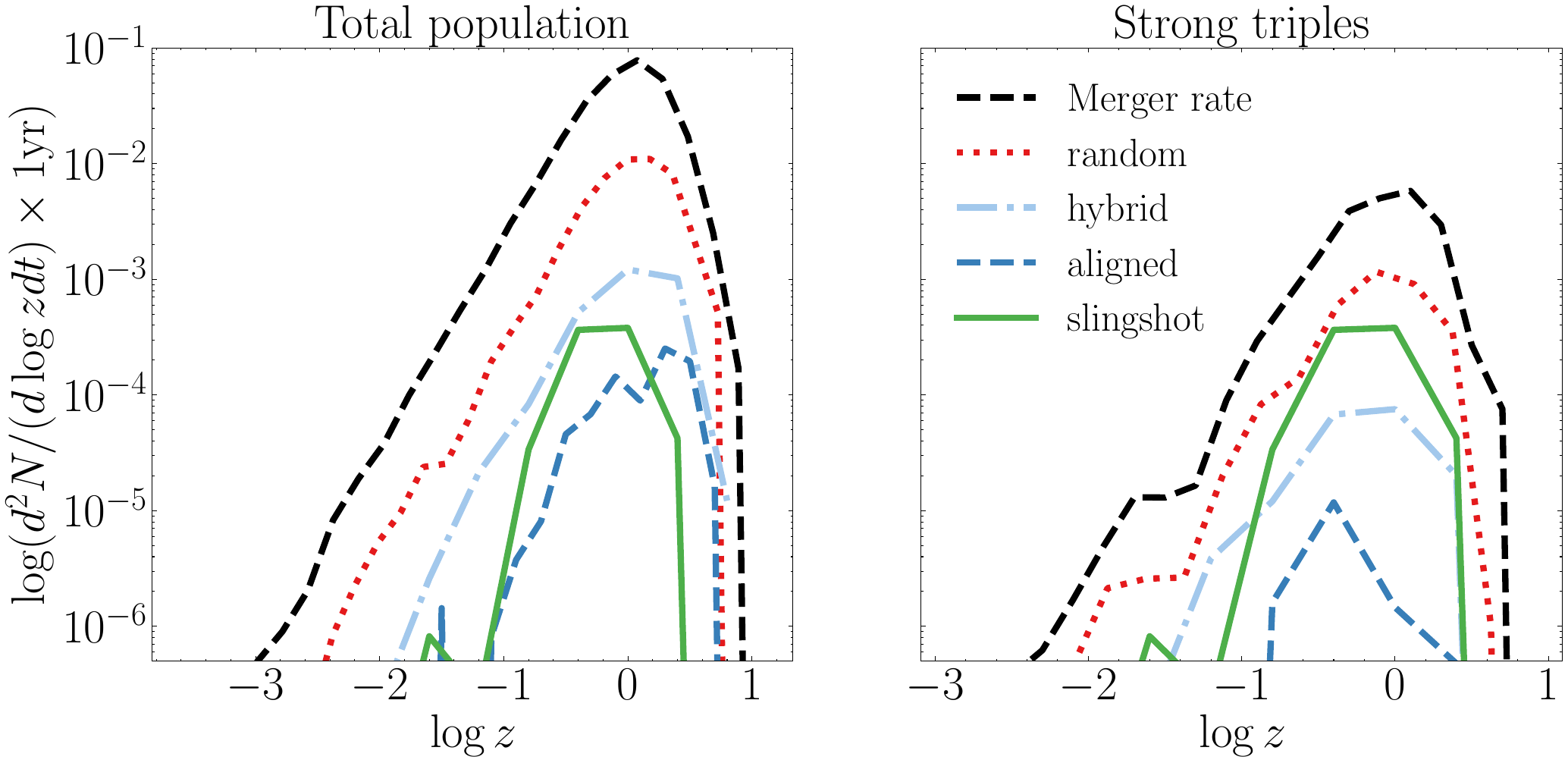}
    \caption{ Ejection rates versus redshift from GW recoil kicks are shown for each spin model: \texttt{random} (red-dotted), \texttt{hybrid} (light blue, dot-dashed) and \texttt{aligned} (dark blue, dashed) spins Also shown is the ejection rate
    from slingshot kicks (solid green), compared with the total {\em merger} rate (black dashed). The left panel shows the ejection and merger rates for the total MBHB population, while the right panel shows only the subpopulation of strong triples.
    The ejection rates are averaged over all realizations. GW recoil from random spins dominate the ejections in the total population contributing to $\sim 7 \%$ of the total MBH population being ejected. Within strong triples, slingshot kick become significant with $\sim 7 \%$ of them resulting in an ejection. }
    \label{fig:escape-and-merger-rates-across-z}
\end{figure*}

We also assess the chances of kicks exceeding the host's central escape velocity, resulting in potential ejections. Around 16\% of GW recoil kicks from \texttt{random} spins exceed
the escape velocity, while 7\% of slingshot kicks surpass this threshold. 
The \texttt{hybrid} spin model results in only 2\% of kicks exceeding escape velocities, while the \texttt{aligned} spin model almost never ejects a BH from its host. 

We now quantify the rates of ejections we see in our population due to the kicks produced from GW recoil and slingshot events. Recall that we are defining ejection for the case when $v_{\text{kick}} > v_{\text{esc}}$, where $v_{\text{esc}}$ is the central escape velocity of the host galaxy. The median escape speed in the total population is $\sim 1269$ \kms. We see from Figure \ref{fig:kick-distributions-for-GW-recoil-and-slingshot} that GW recoil for random spins and slingshot kicks dominate above this range, suggesting they might be efficient in producing ejections. 

However, the total number of ejections from slingshot kicks is limited by the rarity of strong triple interactions. We examine the ejection rates within the population for GW recoil kicks across the \texttt{random}, \texttt{hybrid}, and \texttt{aligned} spin models, as well as for slingshot kicks. We compute the BH ejections through slingshot and gravitational wave recoil as discussed in Section \ref{sec: BH ejections} and calculate the new differential ejection rate using equation \ref{eq: merger rate} where N now is the number of ejections. 

In Figure \ref{fig:escape-and-merger-rates-across-z}, we plot the total ejection rates  from the random, hybrid, aligned GW kicks and slingshot kicks for the total population and the strong triples across all redshifts. Assuming random spins for the total population, the ejection rates 
are approximately an order of magnitude lower than the merger rates (black dashed lines), while slingshot ejection rates are lower by 1–2 orders of magnitude. However, within the strong triple subpopulation, the ejection rates from random GW recoil kicks and slingshot kicks are comparable. In the total population, slingshot kicks produce ejection rates comparable only to GW recoil under hybrid pre-merger spin assumptions.

GW recoil kicks cause ejections of 7.4\%, 0.9\%, and 0.1\% of the total binary population under the \texttt{random}, \texttt{hybrid}, and \texttt{aligned} spin models, respectively, with the \texttt{random} spin model clearly dominating the overall ejections. In contrast, slingshot kicks from strong triple interactions account for 0.4\% of ejections in the total binary population. Within the strong triples subpopulation, slingshot ejections comprises 6.7 \% of ejections. Under the \texttt{random} spin model, GW recoil ejections in strong triples are twice as likely as slingshot ejections. However, if BH spins follow the \texttt{hybrid} model, ejections are five times more likely to be caused by a slingshot kick than by GW recoil.

Ejection rates tend to increase at higher redshifts, primarily due to the combination of lower escape velocities and elevated merger rates. For models such as aligned spins, the ejection rates from GW recoil kicks decrease significantly at lower redshifts. This behavior arises because aligned-spin GW recoil kicks produce a maximum speed of 610 \kms, which becomes insufficient to exceed the larger escape velocities typically found in galaxies at lower redshifts. In contrast, slingshot ejections can generate higher-velocity kicks, maintain the ability to produce ejections even at lower redshifts, where the escape speeds are higher.

MBH ejections will affect future mergers and the total merger rate. Our analysis shows that 5.6\% fewer mergers occur when accounting for ejected BHs that cannot participate in subsequent mergers, for slingshot ejections and GW recoil ejections with the random spin model, of which 9 \% of ejections are via slingshot kicks. In contrast, only 1.6 \% of mergers are impacted by GW recoil ejections in the hybrid spin model, and just 0.2 \% are affected in the aligned spin model. This indicates that ejections have a minimal to modest effect on the merger rate, depending on the spin model. However, this is subject to the caveat that Illustris does not include high-redshift, low-mass seed formation, which could greatly increase the high-redshift merger rate and hence the ejection rate \citep{volonteri_gravitational_2007,Sesana_2007,dunn_role_2020,dongpaez2024,bhowmick2024}. We also calculate the cumulative merger rate, removing the mergers that won't happen due to ejections. From GW recoil kick in random spins and slingshot kicks, the cumulative merger rate decreases from 0.402 yr$^{-1}$ to 0.388 yr$^{-1}$. The ejections from hybrid and aligned spin model produce negligible amount of difference in the merger rate. Again, these merger rates are subject to the same caveats about the limitations of Illustris in the low-mass, high-redshift regime.

\section{Summary and Discussion}

\label{sec: discussion}

In this paper, we have explored strong triple MBH interactions in the Illustris cosmological simulation, by incorporating the results from triple MBH dynamics simulations \citep{bonetti_post-newtonian_2018-1} into the population of strong triples identified via a binary inspiral model \citep{Kelley_2017a, sayeb_mbh_2023}. The fiducial model considers binary hardening due to dynamical friction, stellar scattering, CD and GWs, and assumes a full loss cone. Strong triple interactions cause prompt merger of a pair of MBHs in the triple system or can eject the lightest MBH from the system via gravitational slingshot. We also considered the subsequent evolution of the remaining binary after a slingshot event, to merge via inspiral evolution. Additionally, we investigated the relative occurrence of GW recoil from binary mergers and slingshot kicks in triple systems, both of which can eject an MBH from its host galaxy, creating offsets and wandering BHs. For binary MBH mergers, we calculate GW recoil kick magnitudes for different BH spin models - \texttt{random}, \texttt{hybrid}, and \texttt{aligned}. Circumbinary gas disk produces aligned spins and they produce low-magnitude kicks. The hybrid model incorporates host galaxy gas fractions to assign nearly aligned spins in gas-rich mergers. We compared these GW recoil kicks across spin models with slingshot kick velocities from strong triples. For kicks exceeding the host galaxy's escape speed, the MBH was considered ejected. Our key findings are summarized below:

\begin{itemize}

    \item In our fiducial model,  strong triple interactions increase the overall merger fraction by 4\% (i.e., the fraction of systems that merge by $z=0$). Within the strong triple subpopulation, strong triple dynamics increase the merger fraction from 40\% (binary inspiral alone) to 69\% (after including triple interactions). In a pessimistic model where all isolated binaries stall, triple interactions produce a merger rate that is suppressed by a factor of $4.6$ compared to the fiducial model. Thus, in agreement with \citet{bonetti_post-newtonian_2018}, we find that triple interactions can still produce a GW population observable by PTAs even if other environmental effects fail to do so.

    \item The median mass ratio of mergers in strong triples is $0.63$, which is much greater than the median mass ratio in isolated binary mergers  ($0.25$). We also find that $64\%$ of strong triple mergers are major $q_{\rm mrg} > 0.1$. Massive, major mergers ($M_{\text{mrg}} > 10^{8} M_{\odot}$, $q_{\text{mrg}} >0.1$) are $> 3$ times more likely to be caused by strong triple interactions (relative to all mergers), highlighting the importance of this channel for MBHs observed by PTAs.

    \item Both three-body slingshot kicks and GW recoil from random spins produce high kick speeds. 16\% of GW recoil kicks from random spins and 10\% of slingshot kicks are above 1000 \kms. In contrast, only about $\sim 1\%$ of GW kicks from the hybrid spin model are above this threshold, while the aligned spin model produces none.

    \item Assuming random pre-merger spins, GW recoil results in the ejection of $\sim 7 \%$ of the total population. Within strong triple systems, GW recoil from random spins leads to twice as many ejections as slingshot kicks. However, under a hybrid spin model, slingshot-induced ejections in strong triples are five times more likely than those caused by GW recoil.

    \item If BH spins are randomly oriented, BH ejections via both GW recoil and slingshot kicks reduce the total number of mergers by 6\%; this is dominated by GW recoil ejections. As a result, the cumulative merger rate decreases from 0.402 yr$^{-1}$ to 0.388 yr$^{-1}$ when these ejections are included. In our hybrid and aligned spin models, BH ejections have negligible effect on the merger rate.

\end{itemize}

Our results show the importance of considering strong triple interactions in merging MBHs, as well as the relative contributions of gravitational slingshot and GW recoil to the overall MBH population.  One important caveat is that we treat strong triple interactions separately from the binary inspiral evolution in strong triples. In reality, hardening due to environmental effects in the binary and strong triple interactions likely occur simultaneously. Another important factor in strong triple systems is the eccentricity of binaries undergoing strong triple interactions. \cite{bonetti_post-newtonian_2018-1} observed that prompt mergers in triple MBH systems can drive binaries to high eccentricities. Therefore, we expect strong triple MBHs to have non-negligible eccentricities. This effect is also reflected in the characteristic strain of the GWB, as shown in \cite{bonetti_post-newtonian_2018}, where the model including triple interactions deviates from the $f^{-2/3}$ power law. In future work, we plan to account for eccentricity in strong triples and develop a more self-consistent treatment of the inner binary evolution and strong triple interactions. 

Because of the limitations of modeling BH formation and evolution in large cosmological simulations like Illustris, our study is limited to MBHs above $10^6 M_{\odot}$ and $z\leq7$. Previous studies \citep{Volonteri_2007, Sesana_2007,dunn_role_2020} have shown that recoil and ejections play a bigger role in the early growth of the seed MBHs. \citet{dunn_role_2020} found that GW recoil can affect the early growth of BH seeds and impact the formation pathways of MBHs.  Recoil events can also displace the MBH from the gas-rich center, potentially stifling MBH growth and disrupting the AGN feedback process \citep{blecha_effects_2008}. This can further affect MBH scaling relations \citep{blecha_recoiling_2011}. \citet{dongpaez2024} included GW recoil in a high-resolution cosmological simulation and show that it can affect the growth of MBHs and create a population of wandering BHs. In future work, we aim to investigate the importance of recoil at higher redshifts for different assembly histories using a high-$z$ simulation.

In summary, our results demonstrate that triple interactions are a key channel for driving MBH mergers, especially for massive and major mergers that are very relevant for PTA observations. Furthermore, the relative contributions of GW recoil and slingshot kicks to the ejected MBH population depends strongly on pre-merger MBH spins. GW recoil dominates when random spins are assumed, but if spins are nearly aligned prior to merger, we find that all ejections result from slingshot kicks. Our findings highlight the importance of including triple interactions, GW recoil and slingshot kicks in MBH evolution models to accurately capture the dynamics and growth of MBHs. 

\section{Acknowledgments}
We thank Matteo Bonetti for providing us data on merger fractions from the triple MBH simulations in \citet{bonetti_post-newtonian_2018-1} and for helpful discussions. LB acknowledges support from the Research Corporation for Science Advancement under Cottrell Scholar Award \#27553 and from the National Science Foundation under award AST-2307171.

\bibliography{main}{}
\bibliographystyle{aasjournal}

\end{document}